\documentclass[aps,prd,twocolumn,groupedaddress]{revtex4-2}
\usepackage{graphicx}
\usepackage{xcolor}
\usepackage{hyperref}

\begin{document}


\title{Dark Matter Axions Modulate Dark State Scintillation}

\author{Noah Bray-Ali}
\affiliation{Science Synergy, Los Angles, California, USA 90045}


\date{\today}

\begin{abstract}
	The decay rate of the thallium-iodine dark state in the popular alkali-halide scintillator NaI(Tl) modulates in phase with the dipole anisotropy of the cosmic microwave background (CMB), it is argued. Axion dark matter particles --- at rest in the local CMB rest-frame --- couple to the static magnetic field created in their rest-frame by the static electric quadrupole field within the tetragonally-distorted thallium-iodine octahedron surrounding each dark state. For axions with rest-mass energy near the roughly 0.5 eV dark state activation energy in NaI(Tl), the resulting near-infrared electric field created by this axion electrodynamic interaction modulates the dark state scintillation lifetime by bringing home the excited hole --- from the four in-plane iodine 5$p$ orbitals that share it --- to the Tl 6$s$ shell where it can re-combine with the excited Tl 6$p$ electron to produce the scintillation light.
	
\end{abstract}

\maketitle
{\em Introduction.\textemdash} In 2021 the combined results of the DAMA/LIBRA experiments, along with the results from the earlier DAMA/NaI experiments, showed a clear annual sinusoidal modulation signal in the single-hit double-coincident bulk scintillation event rate within room-temperature large-mass low-radioactivity crystals of the alkali-halide scintillator material, NaI(Tl), consisting of a small concentration of monovalent thallium ions, ${\rm Tl}^+$, doped into the sodium iodide host \cite{dama2021}. The amplitude of the modulation, $A\approx0.01~{\rm counts/day/kg/keV}$, which has $13.7\sigma$ significance and extends over 22 annual cycles, appears within the window of scintillation event energy from 2 to 6 keV where the value of the cut efficiency, $\epsilon\approx0.65$, falls sharply below one, the value it takes outside this window, for the simple pulse shape analysis developed by the DAMA collaboration to discriminate between the relatively slow dark-state scintillation signal and the much quicker dark-count double-coincident photomultiplier noise \cite{dama-psd2000}. Intriguingly, the phase of  the DAMA annual modulation roughly matches that of the dipole anisotropy in the cosmic microwave background (CMB), caused by the orbital motion of the Earth within the solar system barycenter (SSB) rest-frame, albeit with opposite sign: the CMB dipole anisotropy hits its {\em lowest} value of the year on June 12 around the time when the DAMA scintillation rate is at its {\em peak} \cite{cobe_dmr1990,mccabe2014,planck-dip} (Fig.~1).

\begin{figure}
	\includegraphics[width=0.5\textwidth]{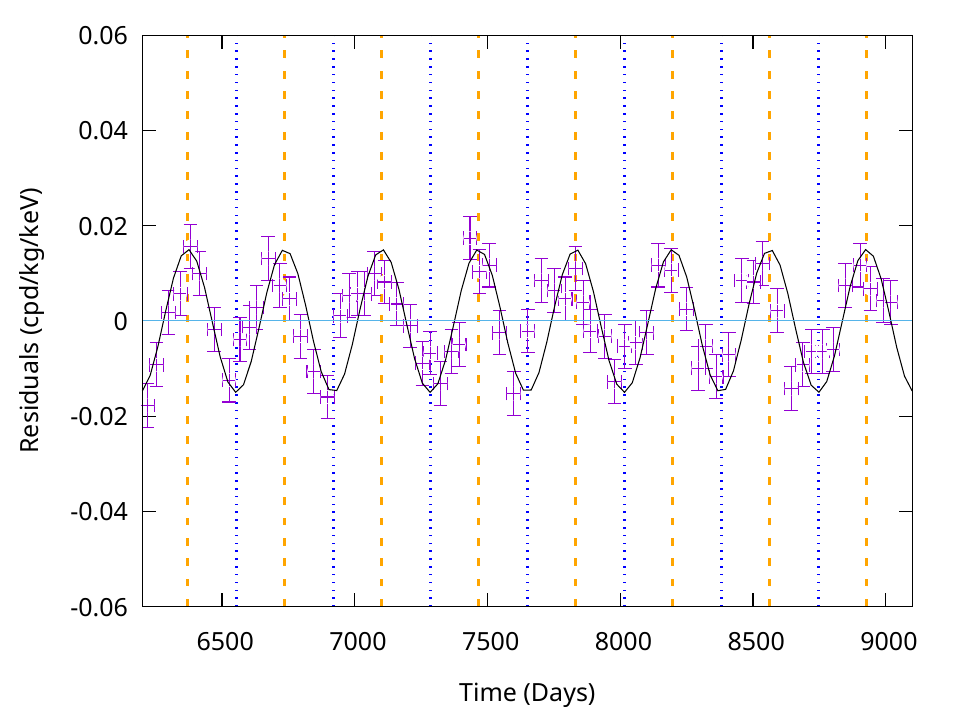}
	\caption{{\bf Dark State Scintillation Modulation} in 8 annual cycles of the DAMA/LIBRA experiment with NaI(Tl) scintillator crystals (data points, \cite{dama2021}) closely follows the predicted modulation due to dark state activation by 0.5 eV QCD axion dark matter at rest in the local cosmic microwave background (CMB) rest-frame (sinusoidal curve). Vertical dashed lines indicate June 12, where the CMB dipole anisotropy reaches its annual minimum, while dotted lines show December 12, where it reaches a maximum.}
\end{figure}
In 2025 a combined analysis of the first three years of data-taking by the COSINE-100 and ANAIS-112 experiments showed no significant annual modulation in the bulk scintillation event rate within NaI(Tl) scintillator crystals with similar mass but significantly more radioactivity compared to those used in the DAMA experiments \cite{anais-cosine2025}. Crucially, both the COSINE and ANAIS teams used a much more sophisticated pulse shape analysis, relying on machine learning, that boosted the cut efficiency to $\epsilon\approx0.95$ throughout the low-energy window from 2 to 6 keV and insured that this high efficiency does not respond significantly to changes in the pulse shape of dark state scintillation events \cite{cosine-psd,anais-psd}. The tension with the DAMA modulation amplitude within this low-energy window, which reached $3.7\sigma$ significance for the combined COSINE-ANAIS results, raises the possibility that the DAMA signal is due to a modulation in the DAMA cut efficiency with amplitude, $\Delta \epsilon/\epsilon\approx0.3\%$, caused by a modulation in the dark state scintillation lifetime, $T$, with size, $\Delta T/T=0.7~\Delta\epsilon/\epsilon\approx0.2\%$ \cite{slow-charge}.

The purpose here is to propose a plausible physical mechanism for a sinusoidal modulation in the dark state lifetime in large-mass low-radioactivity NaI(Tl) crystal scintillators with the period, phase, and amplitude that would explain, in this way, both the dark state scintillation rate modulation observed by the DAMA experiment and the null result reported by COSINE-ANAIS. The key insight is that, cosmologically speaking \cite{khalatnikov1964,lifshitz-landau}, dark matter in the local ``halo'' of the galaxy should be close to rest in the local CMB rest-frame which moves with speed, $v_A=370~{\rm km/sec}$ \cite{cobe_dmr1990,mccabe2014,planck-dip}, relative to the SSB rest-frame. In turn, the sinusoidal modulation of the CMB dipole anisotropy, due to the orbital motion of the Earth in the SSB rest-frame \cite{cobe_dmr1990,mccabe2014,planck-dip}, then implies a corresponding modulation of the NaI(Tl) dark-state velocity in the dark matter rest-frame.

{\em Methods.\textemdash} We assume that dark matter has local energy density, $\rho_A=0.30~(3)~{\rm GeV/cm^3}$ \cite{eilers2019}, and is made of quantum chromodynamic (QCD) axions \cite{wilczek1978,weinberg1978,peccei1989,pdg-axion}, at rest in the local CMB rest-frame, with rest-mass energy, $m_Ac^2\approx0.5~{\rm eV}$, falling close to the dark state activation energy, $\Delta E=0.53~{\rm eV}$, observed in NaI(Tl) crystal scintillators \cite{birks1967,bonanomi1952}. In turn, we interpret this activation energy as an electrostatic charge-transfer energy, $\Delta E=eE_{\rm DC}a$, for moving a hole with positive electric charge, $+e$, from ${\rm I}~5p$ to ${\rm Tl}~6s$, across the Tl-I distance, $a=3.25~{\rm \AA}$ \cite{birks1967}, against the static quadrupole electric field, $E_{\rm DC}=\Delta E/(ea)\approx0.16~{\rm V/\AA}$, created within the Tl-I plane by the tetragonal distortion of the ${\rm TlI_6}$ octahedron around the dark state in NaI(Tl) \cite{color-center1968,seitz1938,mott1940,knox1959,ilingworth1964,haldey1966,delbecq1966}. In the halo axion rest-frame, a static magnetic field then appears within the Tl-I plane with direction that winds spatially with the same phase as a $p_x+ip_y$-wave state for a hole shared among the four in-plane {\rm I} $5p$ orbitals (Fig.~2) \cite{p-wave}, and with strength given by the Lorentz transformation (SI units here and throughout) \cite{lorentz}, 
\begin{equation}
	\label{eq:magnetic-field}
	B_{\rm DC}=(v_A/c)(E_{\rm DC}/c)\overline{\left|\cos\theta_A\right|}=3.3\times10^{-3}~{\rm T},
\end{equation}	
where $\theta_A$ is the angle between the tetragonal axis and the direction of the maximum CMB dipole anisotropy, and the overline indicates the average value, $\overline{|\cos\theta_A|}=1/2$, taken over the three mutually perpendicular crystallographic axes of the cubic NaI crystal, along which the tetragonal axis may lie, and over the daily rotation of the celestial sphere. 
\begin{figure}
	\includegraphics[width=0.5\textwidth]{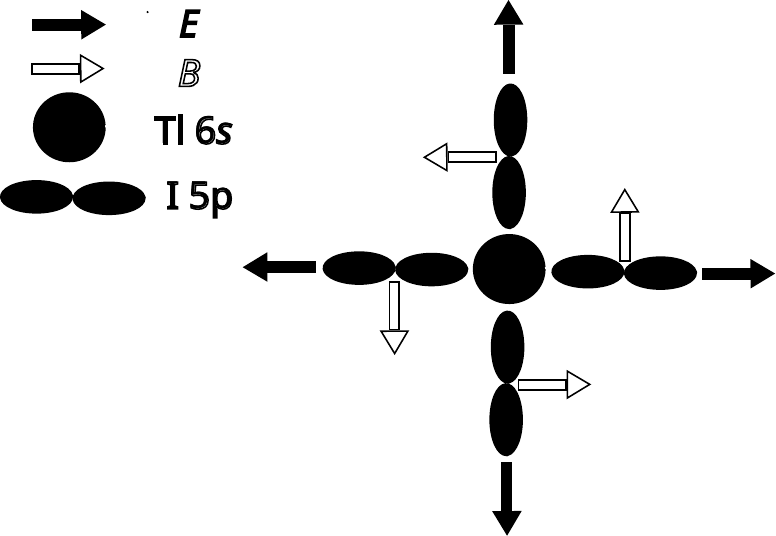}
	\caption{{\bf Thallium-iodine dark state in local cosmic microwave background (CMB) rest-frame}, looking down the tetragonal axis of the ${\rm Tl I_6}$ octahedron on the Tl-I plane with the velocity of the dark state oriented along the axis towards the viewer, shows the four in-plane I 5{\em p} orbitals that share the hole created by charge-transfer from the central Tl 6{\em s} orbital. The static electric field, $E_{\rm DC}$ (filled arrows), created by the Jahn-Teller distortion in the dark state rest-frame, transforms in the local CMB rest-frame to make a small static magnetic field, $B_{\rm DC}$ (unfilled arrows), with strength proportional to the dark state velocity, and with direction that winds spatially with the same phase as a $p_x+ip_y$-wave state for the hole \cite{p-wave}.}
\end{figure}	


Now, in a NaI(Tl) scintillator crystal with volume, $V=2.6\times10^3~{\rm cm}^3$, and dark state lifetime, $T=240~{\rm ns}$, at room temperature, $\theta=292~{\rm K}$, a low-energy bulk scintillation event in the window of interest from 2 to 6 keV creates on average, $N\approx1100$, dark states spread throughout the volume \cite{dama2008,number-dark-states}. Boosting to the halo axion rest-frame, the resulting effective four-dimensional luminous volume, $NVT$, samples from the local halo energy density, $\rho_A$, a Fourier amplitude for the background dynamic axion field, $\tilde{\phi}_A(q_A)$, with four-wavevector, $q_A=(\omega_A,0,0,0)$, set by the axion Compton frequency, $\omega_A=m_Ac^2/\hbar\approx 2\pi\times120~{\rm THz}$. The size of the dark matter halo axion Fourier amplitude scales ``haloscopically'' with the luminous four-volume, $NVT$, as was first noted in the axion dark matter literature more than four decades ago  (Fourier transform convention chosen to make $\tilde{\phi}_A(q_A)$ dimensionless) \cite{sikivie1983,krauss1983}:

\begin{equation}
	\label{eq:axion-fourier}
	\tilde{\phi}_A(q_A)=\sqrt{\rho_ANVT/\hbar}=5.6\times10^{11}.
\end{equation}

Next, the dark matter halo axion Fourier amplitude, $\tilde{\phi}_A(q_A)$, couples to the static magnetic field, $B_{\rm DC}$, created by the dark state in the ${\rm TlI_6}$ octahedron, to create a near-infrared electric field, $E_{\rm NIR}$, polarized parallel to the magnetic field, and oscillating at the axion Compton frequency, $\omega_A$, with field strength \cite{wilczek1978,weinberg1978,peccei1989,pdg-axion},
\begin{equation}
	\label{eq:near-ir}
	E_{\rm NIR}=(B_{\rm DC}c)(m_Ac^2g_{A\gamma\gamma})\tilde{\phi}_A(q_A)\approx0.02~{\rm V/m},
\end{equation}
where $g_{A\gamma\gamma}\approx0.7\times10^{-10}~{\rm GeV}^{-1}$, is the axion-photon coupling strength for a QCD axion with rest-mass energy, $m_Ac^2\approx0.5~{\rm eV}$, and electromagnetic-to-strong axial anomaly ratio, $E/N=8/3$. Combining this with the expression for the static magnetic field strength, in Eq.~(\ref{eq:magnetic-field}), and the expression for the size of the axion Fourier ampitude, in Eq.~(\ref{eq:axion-fourier}), yields the key relation between the dark-state velocity in the local CMB rest-frame, $v_A$, the volume of the NaI(Tl) crystal where the dark states are created by background radioactivity, $V$, and the near-infrared electric field strength created by the axion-photon coupling,
\begin{equation}
	\label{eq:near-ir-full}
	E_{\rm NIR}=E_{\rm DC}\left|\cos\theta_A\right|(v_A/c)(m_Ac^2g_{A\gamma\gamma})\sqrt{\rho_ANVT/\hbar}.
\end{equation}
Clearly, the sinusoidal modulation in the dark state velocity in the dark matter rest-frame with amplitude, $\Delta v_A=v_{\oplus}\approx 30~{\rm km/sec}$, due to the orbital motion of the Earth within the SSB, then translates directly into a corresponding sinusoidal modulation in this near-infrared electric field strength with annual modulation amplitude, $\Delta E_{\rm NIR}$, given by,
\begin{equation}
	\label{eq:near-ir-modulation}
	\frac{\Delta E_{\rm NIR}}{E_{\rm NIR}}=\frac{\Delta v_A}{v_A}=8.1\%
\end{equation}

Finally, the proposed physical mechanism, then, for the modulation in the dark state lifetime is simply that this modulating near-infrared electric field, $E_{\rm NIR}$, interferes constructively with the thermally excited black-body electric field, $E_{\rm BB}=1.1~{\rm V/m}$, with frequency, $\omega=\Delta E/\hbar$, created within the NaI(Tl) scintillator crystal at room temperature, $\theta=292~{\rm K}$ \cite{dama2008}, by the black-body photon energy density, $u(\omega)$, within the linewidth, $\Delta \omega\approx0.24~\omega$ \cite{line-width}, of the near-infrared dark-state activation resonance at photon energy, $\Delta E=0.53~{\rm eV}$ \cite{birks1967, bonanomi1952}, to bring home the excited hole --- from the four in-plane I $5p$ orbitals that share the hole in a $p_x-ip_y$-wave state with phase that winds within the Tl-I plane in the {\em opposite} way as the winding of $E_{\rm NIR}$ created by that of $B_{\rm DC}$ shown in Fig.~2 \cite{p-wave} --- to the Tl $6s$ shell where the hole can re-combine with the excited Tl $6p$ electron to produce the scintillation light \cite{scintillate},
\begin{equation}
	\label{eq:black-body}
	u(\omega)\Delta\omega=\epsilon(\omega) E_{\rm BB}^2\approx\frac{\hbar\Delta\omega}{\pi^2}\left(\frac{\Delta E}{\hbar c}\right)^3\exp(-\Delta E/k\theta),
\end{equation}
where $\epsilon(\omega)\approx3.0~\epsilon_0$ \cite{sodium-thallium-near-ir,polyanskiy2024}, is the dielectric strength of NaI at the near-infrared frequency, $\omega\approx2\pi\times130~{\rm THz}$. The amplitude of the sinusoidal annual dark state scintillation lifetime modulation, $\Delta T$, can then be expressed in terms of the corresponding amplitude, $\Delta E_{\rm NIR}$, given in Eq.~(\ref{eq:near-ir-modulation}), by the interference relation,
\begin{equation}
	\label{eq:lifetime-modulation}
	\frac{\Delta T}{T}=2\frac{\Delta E_{\rm NIR}}{E_{\rm NIR}}\frac{E_{\rm NIR}}{E_{\rm BB}}\approx0.3\%.
\end{equation}
In this way, the sinusoidal annual modulation amplitude, $A$, for single-hit double-coincident bulk scintillation events in NaI(Tl) scintillator crystals in the 2 to 6 keV window with background radioactive event rate, $S=3~{\rm counts/day/kg/keV}$, is determined by this lifetime modulation, for the simple pulse shape analysis done by the DAMA experiments \cite{slow-charge},
\begin{equation}
	\label{eq:event-modulation}
	A=1.4~\frac{\Delta T}{T}S\approx0.01~{\rm counts/day/kg/keV},
\end{equation}
where the phase of the modulation is such that the peak event rate coincides with the maximum cut efficiency, $\epsilon$, which, in turn, corresponds to the {\em longest} dark state lifetime \cite{slow-charge}, and, thus, to the {\em smallest} dark state velocity in the dark matter halo axion rest-frame: June 12 for axions at rest in the local CMB rest-frame \cite{cobe_dmr1990,mccabe2014,planck-dip}.

{\em Results.\textemdash} The resulting modulation in background-radioactivity bulk-scintillation events meets the six criteria established by the DAMA experiment for explaining the observed signal \cite{dama2021}: i) sinusoidal modulation in event rate, ii) having period one year, iii) peaking in early June, iv) confined to the low-energy window from 2 to 6 keV, v) for single-hit double-coincident events, and vi) with modulation amplitude around $A\approx0.01~{\rm counts/day/kg/keV}$. Figure 1 shows the comparison between the predicted sinusoidal modulation and the DAMA data. We emphasize that the predicted sinusoidal modulation has been calculated from first principles with no fit parameters.

{\em Discussion.\textemdash} Having proposed a plausible physical mechanism for the DAMA modulation signal and shown that it meets the stringent criteria put forward by the DAMA collaboration itself for the interpretation of their measurements, we now discuss a wide range of experimental, theoretical, and conceptual challenges to the proposed mechanism. The natural place to begin the discussion is with the regular calibration measurements of the cut efficiency, $\epsilon$, that took place during the DAMA experiment every 10 days or so, and that would seem to imply a modulation of the DAMA signal less than 100 times smaller than the observed result, $A\approx0.01~{\rm cpd/kg/keV}$, if the calibration measurements accurately reflect the value of $\epsilon$ in the NaI(Tl) crystals of the main experiment \cite{dama2008,dama2021}. We show that this null result in the calibration measurements is a straightforward prediction within the framework of analysis leading to the plausible interpretation of the main DAMA experiment results.

The key point is the ``haloscopic'' scaling of the dark matter axion Fourier amplitude, $\tilde{\phi}_A(q_A)\propto\sqrt{V}$, with the luminous volume, $V$, probed in the scintillation measurement, as shown in Eq.~(\ref{eq:axion-fourier}) \cite{sikivie1983,krauss1983}. The standard NaI(Tl) scintillator crystals used in calibration measurements have a much smaller cross-sectional area, typically down by a factor of more than 100, compared to the large production crystals used in the DAMA experiment, and, similarly, the standard calibration high-energy gamma-ray sources, Am-241 59.5 keV and Cs-137 662 keV, have an attenuation length within NaI that is more than 100 times smaller than the thickness of the large production crystals \cite{birks1967}. Combining these two geometric factors with the haloscopic scaling of the modulation amplitude, $A\propto\tilde{\phi}_A(q_A)\propto\sqrt{V}$, it is then straightforward to see that a {\em null} result is to be expected in calibration measurements with the standard small-mass NaI(Tl) scintillator crystals and the Am-241 and Cs-137 sources, since the predicted modulation amplitude for the calibration measurements, $A<\sqrt{10^{-2}\times10^{-2}}~0.01~{\rm cpd/kg/keV}<10^{-4}~{\rm cpd/keV/kg}$, is more than 100 times smaller than the DAMA modulation.

Given the focus on the pulse shape analysis in this discussion, it is important to point out that the first reports from ANAIS used a very similar technique as DAMA, yet already achieved cut efficiency $\epsilon\ge0.8$ throughout the 2 to 6 keV region interest, that was significantly higher than DAMA, and, hence, was less responsive to the small changes in pulse shape that, we argue, are responsible for the DAMA modulation signal \cite{anais-early}. Interestingly, DAMA itself did not use any pulse shape discrimination in its earliest reports \cite{dama-nopsd,dama-nopsd-two}. However, in both of these cases, the DAMA modulation result and also the ANAIS null result, the significance of the results only emerged after the change in pulse shape analysis to one that is sensitive to the dark state scintillation lifetime modulation, in the DAMA case, and to an analysis that is even more ``blind-by-design'' to such modulations than the original pulse shape analysis, in the ANAIS-COSINE case \cite{cosine-psd,anais-psd}.

The discussion of the high-energy gamma-ray DAMA calibration measurements, which probe a small luminous volume and thus suppress the modulation signal, raises an interesting point regarding the low-energy radioactive background events within the production NaI(Tl) scintillator crystals. All three collaborations, DAMA, COSINE, and ANAIS, went to great lengths to suppress the main radioactive background in the 2 to 6 keV low-energy region: 3.2 keV soft x-ray and Meitner-Auger electrons released in the nuclear gamma cascade following electron-capture by the rare radioactive potassium isotope, K-40 \cite{dama2008,anais-cosine2025}. Yet, within the framework of analysis leading to the plausible interpretation of the DAMA modulation result presented here, it is precisely these low-energy radioactive-background bulk scintillation events, such as K-40 3.2 keV, which give rise to the modulation signal by illuminating the full scintillator volume.

Interestingly, the alkali halide scintillator material KI(Tl), with potassium in place of sodium, has a similar dark state thermal activation energy, $\Delta E=0.51~{\rm eV}$ \cite{birks1967,bonanomi1952}, with similar linewidth, $\Delta\omega=0.24~\omega$ \cite{line-width}, and similar dielectric strength at the activation frequency, $\epsilon(\omega)=2.7~\epsilon_0$ \cite{sodium-thallium-near-ir,polyanskiy2024}, as the dark state in NaI(Tl), so that the dark matter activation of the dark state scintillation by 0.5 eV QCD axions at rest in the local CMB rest-frame, discussed so far in the context of NaI(Tl), should also occur in KI(Tl). However, the natural abundance of the radioactive isotope K-40, roughly 0.01\% \cite{dama2008}, is around $10^8$ times larger than the part per trillion K-40 concentrations achieved by DAMA, and approached by COSINE and ANAIS \cite{dama2008,anais-cosine2025}. In a few minutes of measuring a KI(Tl) crystal scintillator, then, a large number of low-energy bulk scintillation events from K-40 3.2 keV can be recorded, roughly $10^4$ to $10^5$~events/kg/keV, that is comparable to the number of events in typical calibration-run measurements of the cut efficiency $\epsilon$ in NaI(Tl) \cite{dama2008}.

Continuing the discussion of a potential measurement of the dark state lifetime modulation using K-40 3.2 keV bulk scintillations in a KI(Tl) scintillator crystal, it is interesting to note that the room-temperature dark state lifetime, $T\approx300~{\rm \mu s}$, is roughly a thousand times larger than that of NaI(Tl) \cite{birks1967,bonanomi1952}. This is relevant because the dark state lifetime modulation amplitude scales ``haloscopically'' not with the luminous volume, $V$, but instead with the luminous four-volume, $VT$, as indicated in Eq.~(\ref{eq:axion-fourier}). Hence, a standard-sized, $5.1~{\rm cm}\times5.1~{\rm cm}\times 5.1~{\rm cm}$, scintillator crystal made of KI(Tl), with twenty times smaller volume than the large-mass production run NaI(Tl) scintillators in the DAMA experiment, would nevertheless yield a roughly 4 times {\em larger} dark-state lifetime modulation amplitude, $\Delta T/T\approx1.1\%$ \cite{potassium}, that is 5 times larger than the noise floor for a calibration-style measurement, with $10^4$ to $10^5$ events/kg/keV per exposure and with temperature control within $\pm0.1~{\rm K}$ around $\theta=292~{\rm K}$ \cite{calibration}, carried out every couple hours over the course of 5 weeks around March 12 or September 12, when the annual CMB dipole anisotropy orbital modulation changes at its fastest rate of the year \cite{cobe_dmr1990,mccabe2014,planck-dip}.

Turning now to a discussion of theoretical challenges to the proposed 0.5 eV QCD axion dark matter mechanism for the DAMA modulation signal, perhaps the clearest challenge is posed by the standard cold dark matter cosmology with cosmological constant ($\Lambda{\rm CDM}$) \cite{planck2018}. Within the framework of $\Lambda{\rm CDM}$, a robust prediction is that the primordial helium abundance produced during Big Bang nucleosynthesis (BBN) within the first three minutes, $Y_P({\rm \Lambda CDM})\approx 0.25$ \cite{primat,parthenope}, should be about ten percent less than the core helium abundance in the Sun, $Y_{\odot}\approx0.28$, when it formed, some 9 billion years later \cite{basu,piersanti}. However, observations of the relative abundance of the helium-burning ``horizontal branch'' stars in globular clusters, which formed in the first billion years after the Bang, suggest a primordial helium abundance, $Y_{\rm HB}\approx0.27$, that is significantly higher than $Y_P({\rm \Lambda CDM})$, when the core-cooling effect of spontaneous 0.5 eV QCD axion dark matter emission --- from the axion-photon coupling to gamma-rays catalyzed by the electric fields of the helium nuclei in the stellar core --- is taken into account in the simulations of stellar evolution that are used to interpret the observations \cite{ayala2014}.

Similarly, the best present observations of the current expansion rate of the universe, by the SH0ES collaboration, have a large tension, now with $7\sigma$ significance, when compared to the value predicted within $\Lambda {\rm CDM}$ \cite{shoes}. A straightforward way to resolve both tensions --- within the same general cosmological framework provided by a flat, homogeneous, isotropic expanding-universe solution to the field equations of general relativity with cosmological constant in which dark matter in the present universe is cold \cite[pg.~100]{peebles} --- would be to double the baryon abundance and cut in half the time it takes to start BBN, while keeping the cosmological constant and the present energy density of cold dark matter roughly the same as in $\Lambda {\rm CDM}$. The main observational constraint on BBN comes from the primordial deuterium abundance \cite{deuterium}, which depends on the product of the baryon abundance with the BBN start time \cite[pgs.~141--143,~184--196]{primat,parthenope,peebles}, and, thus, this constraint would still be satisfied after the proposed changes, which keep this product fixed, yet the primordial helium abundance would rise 10\% or so \cite[pg.~184--196]{primat,parthenope,peebles}, to agree with the value, $Y_P\approx Y_{\rm HB}\approx0.27$, suggested by observations of globular clusters taking into account the core-cooling effect of 0.5 eV QCD axion dark matter on horizontal branch stars \cite{ayala2014}.

Early-universe observations, such as CMB fluctuations, baryon acoustic oscillations, and cosmic shear, have a well-known degeneracy between the baryon abundance and the spectral index of primordial energy density fluctuations \cite{planck2018}. A doubled baryon abundance can, thus, be accommodated by simply increasing this spectral index at the same time. Interestingly, the ${\rm \Lambda CDM}$ value for the spectral index falls significantly {\em below} the value preferred by rather straightforward and remarkably long-standing theoretical arguments \cite{harrison, zeldovich}.

On a more conceptual level, the biggest challenge posed by the proposed mechanism for the DAMA modulation signal is the picture that the axions in the local dark matter ``halo'' of the galaxy are at rest in the local CMB rest-frame. Yet, in the context of baryonic matter structures in galaxies, such as spiral arms, the concept of a quasi-stationary density wave that ``moves through'' the baryonic matter is long established \cite{lin-shu-two,binney}. In the case of the dark matter halo, the corresponding quasi-stationary density wave picture entails that axions at rest in the local CMB rest-frame enter the halo on the Virgo side and leave on the Pisces side, while leaving constant the overall halo density in the galaxy rest-frame \cite{cobe_dmr1990,mccabe2014,planck-dip}.

A similar conceptual challenge is posed by the velocity dispersion of the 0.5 eV QCD axion dark matter in the context of recent null results searching for a sharp mid-infrared emission line coming from the spontaneous two-photon decay of axions in the Milky Way dark matter halo that are in gravitational equilibrium with baryonic matter \cite{pinetti}. Within the framework of analysis used to interpret the DAMA modulation signal, it is more natural, however, to take the axions to be in thermal equilibrium at the CMB temperature, $\theta_{\gamma}=2.73~{\rm K}$ \cite{fixsen}, cosmologically, and to be even warmer, $\theta_A\approx 170~{\rm K}$ \cite{local-halo-temp}, in the local halo. But, this leads to a much broader velocity dispersion, roughly on the order of the ``smooth astrophysical background'' that is removed from the observations \cite{velocity-dispersion}, explaining the null result in \cite{pinetti}.


A rather puzzling challenge, clearly of a conceptual nature, to the 0.5 eV QCD axion dark matter that is invoked in the explanation of the DAMA modulation signal, comes from the interpretation of the neutrino pulse from supernova 1987A. A straightforward analysis of axion emission and absorption within the dense nuclear material in the collapsing core of the supernova progenitor star suggests that QCD axions with rest-mass around 0.5 eV totally dominate the flow of energy and momentum leading to the neutrino pulse \cite{turner}. Although this straightforward result has been interpreted in the supernova literature as an ``exclusion'' of 0.5 eV QCD axion dark matter, we nevertheless are lead by the plausible explanation of the DAMA modulation signal to instead interpret it in a much simpler fashion: the effects of 0.5 eV QCD axion dark matter should be included in the analysis of core collapse supernovae such as 1987A.





{\em Conclusion.\textemdash} The long-standing DAMA sinusoidal annual modulation signal in single-hit double-coincident  bulk scintillation events within large-mass low-radioactivity thallium-doped sodium iodide crystal scintillators in the low-energy window from 2 to 6 keV was analyzed. The analysis offered a plausible physical mechanism for the DAMA modulation signal in terms of the velocity-dependent activation of the scintillating thallium-iodine dark state in the crystal by dark matter in the form of quantum chromodynamic (QCD) axions with rest-mass energy close to the 0.5 eV thermal activation energy for scintillations. Provided the 0.5 eV QCD axion dark matter is at rest in the local cosmic microwave background rest-frame, the resulting modulation in the dark state scintillation lifetime leads, within the simple pulse shape analysis used by the DAMA experiment but not the more sophisticated approach taken by the ANAIS and COSINE experiments, to a sinusoidal annual modulation of the bulk scintillation event rate within the low-energy window that has the roughly the right phase, amplitude, and crystal-volume dependence to explain the DAMA modulation signal and the ANAIS-COSINE null result. 

{\em Acknowledgments.\textemdash} Rita Bernabei provided patient and precise guidance through the published DAMA literature. Ned Wright confirmed that the phase of the orbital modulation of the CMB dipole anisotropy roughly aligns with the DAMA annual modulation signal, triggering the present investigation. The organizers of the 20th Patras Workshop accepted a preliminary version of the present analysis for inclusion in the program.

Daria Mazura, Carol Miu, and Peter John Moraga listened with interest, asked thoughtful questions, and made insightful comments throughout the development of the present work.



\begin{thebibliography}{99}
	\bibitem{dama2021} R. Bernabei et al., Nucl. Phys. At. Energy {\bf 22}, 329 (2021).
	\bibitem{dama-psd2000} R. Bernabei et al., Il Nuovo Cim. A {\bf 112}, 545 (1999).
	\bibitem{cobe_dmr1990} C.~L. Bennett {\em et al.}, Astrophys. J. {\bf 391}, 466 (1992).
	\bibitem{mccabe2014} C. McCabe, J. Cosmol. Astropart. Phys. {\bf 2014}, 027 (2014).
	\bibitem{planck-dip} N. Aghanim et al., Astron. Astrophys. {\bf 641}, A3 (2020).
	\bibitem{anais-cosine2025} N. Carlin et al., Phys. Rev. Lett. {\bf 135}, 121002 (2025).
	\bibitem{cosine-psd} G. Adhikari et al., J. Cosmol. Astropart. Phys. {\bf 2021}, 102581 (2021).%
	\bibitem{anais-psd} I. Coarasa et al., J. Cosmol. Astropart. Phys. {\bf 2022}, 048 (2022). See Figure 10 and accompanying discussion in section 4.3 Results.
	\bibitem{slow-charge} Within the 2 to 6 keV window, DAMA reports roughly $S=3~{\rm counts/day/kg/keV}$ \cite{dama2008}. For a simple exponential decay pulse shape, the mean slow-charge ratio, $\overline{X}_1$, used by the DAMA collaboration for pulse shape discrimination modulates with amplitude, $\Delta \overline{X}_1/\overline{X}_1=0.34~\Delta T/T,$ when the dark-state life-time modulates \cite{dama2008}. Similarly, for the normal distribution of the slow-charge, $X_1$, with mean, $\overline{X}_1=0.7$, and standard deviation, $\sigma=0.1$, reported by DAMA in the 2 to 6 keV window, the cut efficiency modulates with amplitude, $\Delta \epsilon/\epsilon=4.2~\Delta\overline{X}_1/\overline{X}_1=1.4~\Delta T/T$ \cite{dama2008}. The amplitude of the modulation in the rate of bulk radioactive events with slow-charge above the cut is then, $A=(\Delta \epsilon/\epsilon)~S=1.4~(\Delta T/T)~S.$
	\bibitem{khalatnikov1964} A. Khalatnikov and E. Lifshiftz, Adv. in Physics {\bf 12}, 208 (1963). See Section 8.
	\bibitem{lifshitz-landau} E.~M. Lifshitz and L.~D. Landau, {\em The Classical Theory of Fields}, 4th revised English ed., Butterworth Heinemann, London (1975), pgs. 376--381.
	\bibitem{eilers2019} A.-C. Eilers et al., Astrophys. J. {\bf 871}, 120 (2019).
	
	\bibitem{wilczek1978} F. Wilczek, Phys. Rev. Lett. {\bf 40}, 279 (1978).
	\bibitem{weinberg1978} S. Weinberg, Phys. Rev. Lett. {\bf 40}, 223 (1978).
	\bibitem{peccei1989} R.~D. Peccei in {\em CP Violation}, C. Jarlskog ed., World Scientific, Singapore (1989), pgs.~501--551.
	\bibitem{pdg-axion} C.~A.~J. O'Hare et al., {\em Axions and Other Similar Particles} in F. Takahashi et al., Int. J. Mod. Phys. A {\bf 41}, 2630011 (2026).
	
	\bibitem{birks1967} J.~B. Birks, {\em The Theory and Practice of Scintillation Counting}, Pergamon, London (1967).
	\bibitem{bonanomi1952} J. Bonanomi and J. Rossel, Helv. Phys. Acta. {\bf 25}, 725 (1952).
	\bibitem{color-center1968} W. Beall Fowler in {\em Physics of Color Centers}, W. Beall Fowler ed., Academic, New York (1968), pgs.~139, 145--146.
	\bibitem{seitz1938} F. Seitz, J. Chem. Phys. {\bf 6}, 150 (1938).
	\bibitem{mott1940} N.~F. Mott and R.~W. Gurney, {\em Electronic Processes in Ionic Crystals}, Oxford Univ. Press, London and New York (reprinted by Dover, New York, 1964).
	\bibitem{knox1959} R.~S. Knox, Phys. Rev. {\bf 115}, 1095 (1959).
	\bibitem{ilingworth1964} R. Ilingworth, Phys. Rev. {\bf 136}, A508 (1964).
	\bibitem{haldey1966} W.~B. Hadley et al., J. Chem. Phys. {\bf 45}, 2040 (1966).
	\bibitem{delbecq1966} C.~J. Delbecq et al., Phys. Rev. {\bf 151}, 599 (1966).
	\bibitem{p-wave} Taking as the basis states for the hole wavefunction the four in-plane I $5p$ orbitals, $\left|+\hat{x}\right\rangle, \left|+\hat{y}\right\rangle, \left|-\hat{x}\right\rangle,$ and $\left|-\hat{y}\right\rangle,$ labeled by the displacement unit vectors, $\pm\hat{x}, \pm\hat{y}$, from the central Tl site along the pair of crystallographic axes orthogonal to the tetragonal axis (See Fig.~2), the $p_x\pm ip_y$-wave states, $|\pm\rangle$, take the simple form, $\left|\pm\right\rangle=(1/2)(\left|+\hat{x}\right\rangle\pm i\left|+\hat{y}\right\rangle-\left|-\hat{x}\right\rangle\mp i\left|-\hat{y}\right\rangle)$, with eigenvalues, $\pm i$, under the four-fold counter-clockwise rotational symmetry of the ${\rm TlI_6}$ octahedron about the tetragonal axis: $\left|+\hat{x}\right\rangle\rightarrow\left|-\hat{y}\right\rangle\rightarrow\left|-\hat{x}\right\rangle\rightarrow\left|+\hat{y}\right\rangle\rightarrow\left|+\hat{x}\right\rangle$. 
	\bibitem{lorentz} A. Einstein, Ann. Phys. (Leipzig) {\bf 322}, 891 (1905).
	\bibitem{number-dark-states} The number of dark states, $N=3.2~{\rm keV}/3.0~{\rm eV}=1.1\times10^{3}$, is estimated for the K-40 3.2 keV soft-x-ray and Auger electron peak that dominates the low-energy 2 to 6 keV spectrum of single-hit double-coincident events in NaI(Tl) for the DAMA experiment \cite{dama2008}. The dark state scintillation photon energy is $hc/(413~{\rm nm})\approx3.0~{\rm eV}$ in NaI(Tl) at room-temperature \cite{dama2008}.
	
	\bibitem{dama2008} R. Bernabei et al., Nucl. Instrum. Meth. A {\bf 592}, 297 (2008).
	\bibitem{sikivie1983} P. Sikivie, Phys. Rev. Lett. {\bf 15}, 1415 (1983).
	\bibitem{krauss1983} L. Krauss et al., Phys. Rev. Lett. {\bf 55}, 1797 (1985).
	\bibitem{anais-early} J. Amar\'{e} et al., Phys. Rev. Lett. {\bf 123}, 031301 (2019).
	\bibitem{dama-nopsd} R. Bernabei et al., Phys. Lett. B {\bf 424}, 297 (1998).
	\bibitem{dama-nopsd-two} R. Bernabeti et al., Phys. Lett. B {\bf 450}, 448 (1999).
	\bibitem{line-width} The line-width, $\Delta\omega=0.24~\omega$, of the near-infrared charge-transfer dark-state activation with photon energy, $\Delta E=\hbar\omega\approx0.53~{\rm eV}$ \cite{birks1967,bonanomi1952}, in NaI(Tl) is estimated from the line-width of the broad near-infrared absorption resonance at photon energy, $\Delta E\approx0.55~{\rm eV}$, in the charge-transfer insulator, ${\rm La_2CuO_4}$, reported by J.~D. Perkins et al., Phys. Rev. Lett. {\bf 71}, 1621 (1993).  
	
	\bibitem{scintillate} Following the standard picture of the ``lattice-relaxed, electronically-excited'', $V_K+{\rm Tl}^0$, dark-state color-center in the thallium-doped alkali halide scintillator literature \cite{color-center1968,seitz1938,mott1940,knox1959,ilingworth1964,haldey1966,delbecq1966}, we imagine the dark state to be a spin-orbit entangled state formed from the $p_x\pm ip_y$-wave states, $|\pm\rangle$ \cite{p-wave}, for the hole on the four in-plane I $5p$ orbitals, and from the Tl~$6p_z$ orbital for the excited electron with $z$-quantization axis for both orbital and spin angular momentum taken along the tetragonal axis in the direction of the dark-state velocity in the local CMB rest-frame (toward the viewer in Fig.~2): $|V_K+{\rm Tl}^0\rangle=(1/\sqrt{2})(\left|\uparrow\uparrow\right\rangle|-\rangle+\left|\downarrow\downarrow\right\rangle\left|+\right\rangle)\left|{\rm Tl}^0~6p_z^1\right\rangle$. Then, in the presence of the near-resonant, spatially-winding, near-infrared electric field, $E_{\rm NIR}$, which transforms in the same way as $\left|+\right\rangle$ under the four-fold rotational symmetry of the ${\rm TlI_6}$ octahedron about its tetragonal axis (See Fig.~2, with $E_{\rm NIR}$ parallel to $B_{\rm DC}$), a symmetry-allowed electric-dipole transition is possible, $V_K+{\rm Tl}^0\rightarrow {\rm Tl}^+$, from the first term in the dark state, $\left|\uparrow\uparrow\right\rangle\left|-\right\rangle\left|{\rm Tl}^0~6p_z^1\right\rangle$, to a component of the ``bright'' state, ${\rm Tl}^+~6s^16p^1~{}^3P_1$, with wavefunction, $\left|\uparrow\uparrow\right\rangle\left|{\rm Tl}^+~6s^1~6p_z^1\right\rangle$. Finally, this bright state scintillates, releasing the visible photon with wavelength $413~{\rm nm}$ seen by the DAMA experiment \cite{dama2008}, and returning to the electronic ground-state, ${\rm Tl}^+~6s^2~ {}^1S_0$, after which the ${\rm TlI_6}$ octahedron returns to the cubic symmetry of the NaI lattice by reversing the Jahn-Teller tetragonal distortion.
	\bibitem{sodium-thallium-near-ir} H.~H. Li, J. Phys. Chem. Ref. Data {\bf 5}, 329 (1976).
	
	\bibitem{polyanskiy2024} M. N. Polyanskiy, Sci. Data {\bf 11}, 94 (2024)
	\bibitem{potassium} For a room-temperature measurement of K-40 3.2 keV scintillation events in KI(Tl) scintillator crystal with dark state number, $N=1.1\times10^3$ \cite{number-dark-states}, crystal volume, $V=(5.1~{\rm cm}^3)=131~{\rm cm^3}$, and dark state lifetime, $T=300~{\rm \mu s}$ \cite{birks1967,bonanomi1952}, the scaling of the axion Fourier amplitude, $\tilde{\phi}(q_A)=5.6\times10^{11}(N/1.1\times10^{3})^{1/2}(V/2.6\times10^3~{\rm cm^3})^{1/2}(T/0.24~{\rm \mu s})^{1/2}$, given in Eq.~(\ref{eq:axion-fourier}), yields an eight-fold enhancement, $\tilde{\phi}_A(q_A)=8.0\times(5.6\times10^{11})=4.5\times10^{12}$, compared to the DAMA case. Meanwhile, the dark state activation energy, $\Delta E=0.51~{\rm eV}$ \cite{birks1967,bonanomi1952}, and Tl-I distance, $a=3.12~{\rm \AA}$ \cite{birks1967}, yield roughly the same value in KI(Tl) as in NaI(Tl) for the static electric quadrupole field, $E_{\rm DC}=\Delta E/(ea)=0.16~{\rm V/\AA}$, so that the eight-fold enhancement passes through to the near-infrared electric field at the axion Compton frequency created by the motion of the dark state within the local CMB rest-frame, $E_{\rm NIR}\approx8.0\times0.02~{\rm V/m}\approx0.16~{\rm V/m}$, using Eq.~(\ref{eq:near-ir}) for the DAMA value. Finally, the black-body electric field strength, $E_{\rm BB}=2.3~{\rm V/m}$, given by Eq.~(\ref{eq:black-body}), has about twice the size as DAMA, due to the slightly smaller activation energy, $\Delta E=\hbar\omega=0.51~{\rm eV}$, similar linewidth $\Delta\omega=0.24~\omega$ \cite{line-width}, and somewhat smaller dielectric strength, $\epsilon(\omega)=2.7~{\rm \epsilon_0}$ \cite{sodium-thallium-near-ir,polyanskiy2024}, in KI(Tl) compared to NaI(Tl), so that the overall bulk scintillation life-time modulation amplitude, $\Delta T/T=0.3\%~(E_{\rm NIR}/0.02~{\rm V/m})~(E_{\rm BB}/1.1~{\rm V/m})^{-1}$, from Eq.~(\ref{eq:lifetime-modulation}), is around four times larger than DAMA: $\Delta T/T\approx1.1\%$.  
	
	\bibitem{calibration} Over the course of 35 days around March 12 or September 12, the room-temperature dark state scintillation lifetime in KI(Tl), $T=300~{\rm \mu s}$ \cite{birks1967,bonanomi1952}, changes by the amount, $\Delta T=T~(\Delta T/T)~(2\pi\times35~{\rm d}/365~{\rm d})\approx2~{\rm \mu s}$, for the sinusoidal dark state lifetime modulation with period one year, amplitude, $\Delta T/T\approx1.1\%$ \cite{potassium}, and maximum around June 12 driven by the annual orbital modulation of the dark-state speed in the local CMB rest-frame \cite{cobe_dmr1990,mccabe2014,planck-dip}. Meanwhile, for a temperature control at room-temperature, $\theta=292~{\rm K}$, within $\sigma_{\theta}=0.1~{\rm K}$, the thermal activation of the dark state lifetime by the black-body electric field with strength, $E_{\rm BB}\propto \exp(-\Delta E/k\theta)$, given in Eq.~(\ref{eq:black-body}), with the dark state thermal activation energy value in KI(Tl), $\Delta E=0.51~{\rm eV}$ \cite{birks1967,bonanomi1952}, gives the dominant systematic uncertainty in each measurement of the dark state lifetime, $\sigma_{T}/T=(\sigma_{\theta}/\theta)~(\Delta E/k\theta)=0.7\%$. Finally, choosing the exposure time of the calibration-style measurement to yield between $10^4$ to $10^5$ counts/keV/kg of K-40 3.2 keV dark state scintillation \cite{dama2008}, gives a matching statistical uncertainty, for an overall uncertainty of each measurement, $\sigma_T=\sqrt{2}\times 0.7\%\times T=3~{\rm \mu s}$, so that the resulting campaign of measurements, performed every couple hours for 35 days, yields an uncertainty in the measurement of the change in dark state scintillation lifetime, $\sigma_{\Delta T}=2\sigma_T\sqrt{2/(35\times13)}=0.4~{\rm \mu s}$, that is 5 times smaller than the predicted change, $\Delta T\approx2~{\rm \mu s}$, for measurement times chosen using the roots of the Chebyshev polynomial of appropriate order, following the recipe for Chebyshev approximation given in W.~H. Press, S.~A. Press, W.~T. Vetterling, and B.~P. Flannery, {\em Numerical Recipes in C}, 2nd ed., Cambridge Univ. Press, Cambridge (1997), pgs. 190--194.      
	
	\bibitem{planck2018} N. Aghanim et al., Astron. Astrophys. {\bf 641}, A6 (2020).
	\bibitem{primat} C. Pitrou, A. Coc, J.-P. Uzan, and E. Bangioni, Phys. Rep. {\bf 754}, 1 (2018).
	\bibitem{parthenope} S. Gariazzo, P.~F. de Salas, O. Pisanti, and R. Consiglio, Comp. Phys. Comm. {\bf 271}, 108205 (2022).
	\bibitem{basu} A.~M. Serenelli and S. Basu, Astrophys. J. {\bf 719}, 865 (2010).
	\bibitem{piersanti} L. Piersanti, O. Straniero, and S. Cristallo, Astron. Astrophys. {\bf 462}, 1051 (2007).
	\bibitem{ayala2014} A. Ayala et al., Phys. Rev. Lett. {\bf 113}, 191302 (2014).
	\bibitem{shoes} A.~G. Riess et al., Astrophys. J. Lett. {\bf 992}, L34 (2025).
	\bibitem{peebles} P.~J.~E. Peebles, {\em Principles of Physical Cosmology}, Princeton Univ. Press, Princeton (1993).
	\bibitem{deuterium} R.~J. Cooke, M. Pettini, and C.~C. Steidel, Astrophys. J. {\bf 855}, 102 (2018).
	\bibitem{harrison} E.~R. Harrison, Phys. Rev. D {\bf 1}, 2726 (1970)
	\bibitem{zeldovich} Y.~B. Zel'dovich, Mon. Not. Roy. Astron. Soc. {\bf 160}, 1P (1972).
	
	\bibitem{lin-shu-two} C.~C. Lin and F.~H. Shu, Proc. Nat. Acad. Sci. {\bf 55}, 229 (1966).
	\bibitem{binney} J. Binney and S. Tremaine, {\em Galactic Dynamics}, Princeton Univ. Press, Princeton (2007), pgs.~525--526.
	
	\bibitem{pinetti} E. Pinetti, arXiv:2503.11753 (2025).
	
	\bibitem{fixsen} D.~J. Fixsen, Astrophys. J. {\bf 707}, 916 (2009).
	
	\bibitem{local-halo-temp} In the local dark matter halo, the number density of 0.5 eV QCD axions, $\rho_A/m_Ac^2\approx 6\times10^8~{\rm cm^{-3}}$ \cite{eilers2019}, is much larger than the cosmological value, $n_A=n_{\gamma}~(\Omega_Ah^2/\Omega_{\gamma}h^2)~(k\theta_{\gamma}/m_Ac^2)~(u_{\gamma}/n_{\gamma}k\theta_{\gamma})\approx6~n_{\gamma}=2.5\times10^3~{\rm cm^{-3}},$ obtained using the best available estimate of the cosmological energy density parameter for cold dark matter in the present universe, $\Omega_Ah^2=0.119$ \cite{planck2018}, and the properties of the CMB black-body distribution at temperature, $\theta_{\gamma}=2.73~{\rm K}$ \cite{fixsen}: number density, $n_{\gamma}=411~{\rm cm^{-3}}$ \cite[pg.~158--159]{peebles}, energy density, $u_{\gamma}=(3!\zeta(4)/2!\zeta(3))~n_{\gamma}k\theta_{\gamma}=2.70~n_{\gamma}k\theta_{\gamma}$ \cite[pg.~137--138,~158--160]{peebles}, and cosmological energy density parameter, $\Omega_{\gamma}h^2=2.47\times10^{-5}$ \cite[pg.~100]{peebles}. To keep fixed the phase-space density, as one must for these basically ``collisionless'' particles \cite{wilczek1978,weinberg1978,peccei1989,pdg-axion}, then, the temperature must rise in the local halo to the value, $\theta_A=\theta_{\gamma}~(\rho_A/n_Am_Ac^2)^{1/3}\approx170~{\rm K}$, using the number density's dependence on temperature at fixed fugacity for a cold, but not condensed, ideal Bose gas from K. Huang, {\em Statistical Mechanics}, 2nd ed., Wiley, New York (1987), pg.~188--189.
	
	\bibitem{velocity-dispersion} The search for the 0.5 eV QCD axion dark matter two-photon decay ``line'' at wavelength, $2\lambda_A=2h/m_Ac\approx5~{\rm \mu m}$, uses a window of width, $\Delta \lambda\approx0.2~{\rm \mu m}$, in which a quartic polynomial is fit to the spectrum and subtracted from the ``blank-sky'' observations of the Milky Way dark matter halo with the MIRI mid-infrared spectrometer aboard the James Webb Space Telescope \cite{pinetti}. However, the velocity dispersion, $\sigma_v=c\sqrt{3k\theta_{\gamma}/m_Ac^2}\approx 11,000~{\rm km/sec}$, for 0.5 eV QCD axions in thermal equilibrium with the CMB at temperature, $\theta_{\gamma}=2.73~{\rm K}$ \cite{fixsen}, leads to a thermal broadening, $2\Delta \lambda_A=2(\sigma_v/c)\lambda_A\approx0.2~{\rm \mu m}$, on the same order as the window width, $\Delta\lambda\approx0.2~{\rm \mu m}$. At the higher temperature, $\theta_A\approx 170~{\rm K}$ \cite{local-halo-temp}, found in the local dark matter halo, which is still ``cold'' in terms of the comparison, $k\theta_A\ll m_Ac^2$, the broadening is even larger.
	
	\bibitem{turner} M.~S. Turner, Phys. Rev. Lett. {\bf 60}, 1797 (1988).
\end{thebibliography}
\end{document}